\documentclass[conference]{IEEEtran}
\usepackage{graphicx}
\usepackage{amsmath,amssymb,epsfig,xcolor} % easier math formulae, align, subequation
\usepackage{subfig}
\usepackage{tabularx,multirow,array}
\usepackage{times}
\usepackage{booktabs}
\usepackage{multirow}
\usepackage{soul}
\usepackage{enumerate}
\usepackage{xcolor}
\usepackage[countmax]{subfloat}
\usepackage{algpseudocode}
\usepackage{algorithm}
\usepackage{threeparttable}
\usepackage{cite}
\usepackage{url}
\usepackage{amsthm}
\usepackage{bbm}
\usepackage{syntax}
\usepackage{newfloat}

\DeclareFloatingEnvironment[
  fileext   = logr,% don't use log here!
  listname  = {List of Grammars},
  name      = Grammar,
  placement = htp
]{Grammar}

\theoremstyle{plain}

\SetMathAlphabet{\mathcal}{bold}{OMS}{cmsy}{b}{n}
\begin{document}
\title{The Role of Intent-Based Networking in ICT Supply Chains}
\author{
\IEEEauthorblockN{Mounir Bensalem, Jasenka Dizdarevi{\'{c}}, Francisco Carpio, and Admela~Jukan}
\IEEEauthorblockA{Technische Universit\"at Braunschweig, Germany}
\IEEEauthorblockA{\{mounir.bensalem, j.dizdarevic, f.carpio, a.jukan\}@tu-bs.de}
}
\maketitle
\begin{abstract}
The evolution towards Industry 4.0 is driving the need for innovative solutions in the area of network management, considering the complex, dynamic and heterogeneous nature of ICT supply chains. To this end, Intent-Based networking (IBN) which is already proven to evolve how network management is driven today, can be implemented as a solution to facilitate the management of large ICT supply chains. In this paper, we first present a comparison of the main architectural components of typical IBN systems and, then, we study the key engineering requirements when integrating IBN with ICT supply chain network systems while considering AI methods. We also propose a general architecture design that enables intent translation of ICT supply chain specifications into lower level policies, to finally show an example of how the access control is performed in a modeled ICT supply chain system.
\end{abstract}

\section{Introduction}

In the recent years intent-based networking (IBN) has been the emerging
networking paradigm, levering the integration of AI/ML techniques with network
orchestration to improve upon the network automation, complexity and control
functionalities. Compared to traditional network management solutions it has
shown great potential in numerous aspects, reducing the time for network
configuration deployment while improving on its scalability\cite{Shry},
increasing both the efficiency and reliability, as well as reducing costs in
particularly complex industrial networks \cite{saha2018intent} and integrating
security through its intent functions \cite{laliberte2018journey}. The key
behind IBN's increasing abilities to solving complex network tasks lies is in
their high-level policies specification capabilities that allow network
administrator to define goals and demands on how the network should behave.

One of the particularly significant IBN potential application domains are ICT
supply chains, since the fast-growing technological evolution and the digital
transformation behind Industry 4.0 and ICT supply chains have created a gap
between traditional networking solutions and these emerging complex networks. As
now ICT supply chains  can span across multiple vendor infrastructures and
devices, assuming different configuration requirements and capabilities with
continuous addition of new functionalities, the interest in novel network
management and orchestration approaches has been growing rapidly. The
utilization of IBN in this domain could facilitate automation, control and
orchestration processes, while improving on the network robustness and operation
dynamicity.

This paper introduces an IBN approach to automating the configuration management
and control of ICT supply chain networks through high-level intents.  The
objective is to first analyse the key engineering requirements in integrating
IBN approach with these complex networks, considering the questions of
centralized vs decentralized orchestration, influence of different AI
techniques, and manner of specifying high level intents to ICT supply chain
policies. To this end we propose an architectural design of intent based
translator for ICT supply chains, as well as an  example in the context of
supply chain assets access control.

The rest of the paper is organized as follows.  Section 2 describes the
background and motivation behind using intent-based approaches. Section 3 gives
an overview of engineering requirements, the general architectural design for
utilization of IBN in ICT supply chains, and a use case scenario for assets
access control. Finally, Section 4 concludes the paper.

\section{Background and motivation}

In this section we analyse the common architectural building blocks of any intent-based system, according to the already existing solutions. While the design of intent-based translators can follow guidelines depending on the scenario to be applied, this analysis was used as the initial step in proposing an IBN architectural design of the ICT supply chain system. Table \ref{tab:components} shows a comparative analysis of the intent abstraction layer design for ICT systems. Several proposals from the literature are investigated to understand the main software components  responsible on interfacing with the user, storage, management, policies, networking, monitoring, and enabling intelligence. It is important to mention that few proposals highlighted the role of AI and ML in translating user intents \cite{mahtout2020using} and improving policies using insight from telemetry \cite{8638149, abbas2020slicing, khan2020intent}.

\begin{table*}[ht]
    \centering
    % \scriptsize
    {\begin{tabular}{p{0.5cm}p{1.5cm}p{2.5cm}p{1.5cm}p{1.5cm}p{1.5cm}p{1.5cm}p{1.5cm}p{2.2cm}}
            \toprule
            \textbf{Work}               & \textbf{Interface} & \textbf{Management}        & \textbf{Storage}                      & \textbf{Policies}           & \textbf{Network}      & \textbf{AI/ML} & \textbf{Monitoring} & \textbf{Application}             \\
            \midrule
            \cite{riftadi2019p4i}       & GUI                & intent parser \& compiler  & service library                       & policy builder              & switch controller     & no             & no                  & switches programmability         \\
            \midrule
            \cite{han2016intent}        & extended NBI       & intent manager \& compiler & vocabulary store                      & conflict checking           & SDN controller        & no             & yes                 & ---                              \\
            \midrule
            \cite{chaudhari2019vivonet} & GUI, Alexa         & intent engine \& compiler  & MariaDB                               &                             & floodlight controller & ---            & no                  & network visualization            \\
            \midrule
            \cite{pham2016sdn}          & UI                 & intent manager \& compiler & service registry                      & service integration element & SDN controller        & no             & no                  & SDN applications                 \\
            \midrule
            \cite{jain2020intent}       & GUI, Alexa         & intent engine              & database                              &                             & Ryu controller~       & no             & ---                 & self-Healing SDN                 \\
            \midrule
            \cite{khan2020intent}       & NBI                & intent manager             & policy store                          & policy configurator         & resource controller   & LSTM, GAN      & yes                 & orchestration of network slicing \\
            \midrule
            \cite{8638149}              & NBI                & policy parser              & policy store                          & conflicts management        & SDN controller        & heuristics     & yes                 & SDN-packet processing            \\
            \midrule
            \cite{subramanya2016intent} & REST API           & intent engine              & no                                    & PCE                         & Ryu Controller        & no             & no                  & mobile backhauling               \\
            \midrule
            \cite{abbas2020slicing}     & GUI                & intent manager             & policy store                          & policy configurators        & FlexRAN, open MANO    & DL, GAN        & yes                 & network slicing                  \\
            \midrule
            \cite{mahtout2020using}     & slack API          & intent manager \& compiler & ontology  and RDF                     & conflict checking           & ---                   & LSTM           & ---                 & high-speed networks              \\
            \midrule
            \cite{chung2020design}      & NETCONF            & extractor \& compiler      & ---                                   & convertor                   & ---                   & no             & no                  & IoT device configuration         \\
            \midrule
            our work                    & GUI                & intent manager \& compiler & policy store, intent store, telemetry & policy configurator         & resource, controller  & RL, LSTM       & yes                 & ICT supply chain                 \\
            \bottomrule
        \end{tabular}}
    \caption{Comparative analysis of intent-based architectural components}
    \label{tab:components}
\end{table*}
\vspace{-0.2 cm}
\subsection{Intent-Based Translator Design}

\subsubsection{High-Level Language}

The definition of a high level intent-based language is essential to allow the
user to express its requirements to the system. Multiple works in the literature
try to define this type of language, for instance, Riftadi et al
\cite{riftadi2019p4i} employed Nile, a network intent language that provides an
intermediate layer between natural language and lower-level policies. In \cite{procera2012}  a high level declarative policy language is proposed
to control the architecture of software-defined networking (SDN) based on the
notion of functional reactive programming where policies are used to state the
change in network conditions. In \cite{8638149}, it is defined an Open Software
Defined Framework (OSDF) policy language used to provide a high level API in
order to express network requirements by managers and  network  administrators.
\cite{scheid2020controlled} uses a Controlled Natural Languages (CNL), which
sets restrictions on  inputs by the definition of grammar model. In this
case, CLN was employed to control blockchain selection queries, as mapping 
requirements to a particular blockchain needs expertise and knowledge about the
existing blockchain implementations. Recent development of natural language
processing (NLP) algorithms \cite{chaudhari2019vivonet, jain2020intent} show how
voice-assisted technologies have been used to manage the network as well.

\subsubsection{Interfaces}

The initial works on intent-based interfaces support imply their abilities of managing internal and external interactions independently from the underlying network technologies. In the approach by Szyrkowiec et al
\cite{szyrkowiec2018automatic}, the intents issued by a client application, are
submitted through a representational state transfer (REST) northbound interface
(NBI). Similarly, in the work by Zeydan et al \cite{zeydan2020recent}, intent
based NBI of NIC allows descriptive way to obtain what is desired from the
network using any kind of protocols, including OpenFlow, BGP, and Netconf.
%Subramanya et al \cite{subramanya2016intent} implemented and validated their proposed intent–based networking interface using the EmPOWER platform. EmPOWER is an open source tool used to for SDN/NFV  research  in wireless and mobile networks. EmPOWER relies on a centralized controller to implement control and management tasks. 
In the work by Pham et al \cite{pham2016sdn}, the controller’s
NBI and core services are used as the atomic layer in the service orchestration
because the application wants to retrieve abstract network topology provided by
the controller. The intent layer in Han et al \cite{han2016intent} provides
extended North Bound Interface (NBI) that is used by various applications.

Mahtout et al \cite{mahtout2020using}, deployed EVIAN, an approach for using ML
for intent-based provisioning, the EVIAN client provides an intelligent chatbot
interface, via Slack, to users in order to understand their network requirements
and science needs. % NLP  techniques are used to extract actions from the intents. Once users agree, EVIAN Renderer maps intent entities into network device actions, via available device directory, to generate network APIs that can automate configurations. Sk\"oldstr\"om et al \cite{skoldstrom2017dismi} chose a REST-based interface due to its ease of use and compatibility most programming languages. Authentication and encryption can be supported through HTTPS, making it easy to implement secure services. The DISMI API has been specified using Swagger, a language for specifying both the methods of the API and data models in JSON or YAML. Based on the specification, different Swagger tools can generate client- and server-side source code for several programming languages as well as documentation. The server-side provides a REST-based API that publishes a Swagger JSON file with the full API specification.

\subsubsection{Manager/Parser}
In the \cite{saha2018intent},  authors have recognized a core component of IBNs, which
interacts with most of the other components, naming it an Intent Processor.
This core component is responsible for accepting high level intents expressed
by users (applications) using a semantic language or through other appropriate
user interfaces. For instance, it can receive input in plain English and, in
turn, it can translate that into say, Algebraic Modelling Language, which can be
easily understood by a mathematical solver. In \cite{8638149}, the similar functionality has been
been identified under the name Policy Parser, responsible for analyzing application-based
policies.

\subsubsection{Policy Builder}

In the work by Abbas et al \cite{abbas2020slicing} on creating core network
slices through IBN, the policy configurator extracts the information related to
resources from the graph, provided by the intent manager and converts the
specifications into a slice template. That slice template is according to the
network orchestrator acceptable format; for example, for the OSM core, the
policy configurator can give the slice template in  form of a JSON file which
contains the information related to the deployment of  network functions and
their mapping.

\subsubsection{Compiler}
The intent compiler is one of the key functionalities related to IBN system design,
with the task of converting the translated specifications or policies into technical configurations
expected to realise the intent. As such it has been a focus of numerous research works.
In \cite{chung2020design} it is used to generate a low-level configuration XML
format from the low-level configuration data using context-free grammar CFG. In
\cite{yang2018automata} extracted data is converted to network security
function required data via Data Converter. In \cite{riftadi2019p4i} an element
called P4 code generator is proposed to generate a working P4 program with the
input of a policy graph introduced by the user as an intent, while in
\cite{skoldstrom2017dismi} generic constraints are converted into
technology-specific ones, to be later applied by network controller.

\subsubsection{Knowledge Base (KB)}
The KB is a database that stores all the necessary information for
the intent based system, such as predefined policies, intent vocabularies and
telemetry used for improving policy building and intent translation process.
In the recent IBN works, KB has been mostly recognized under the name
policy store \cite{khan2020intent, abbas2020slicing, 8638149}. Its main tasks include storing and retrieving
application-based network policies that a user enters to the system. As such,
it has been used to access the current active policies in the system, allowing network
administrators to create, read, update, delete policies at runtime. In systems
like \cite{riftadi2019p4i}, the policy store represents an action library
containing  P4 code templates, that can be parsed by the intent engine into a
policy graph.

\vspace{-0.2 cm}
\subsection{Application in ICT systems}

With an increased interest in the automation of network orchestration, the emerging
IBN approaches are being extensively analysed for its utilization in numerous
application fields, from vehicular, IoT and 5G scenarios, to its  potential in
provisioning intent-based security policies and  services.

One of the most interesting application assumed to benefit of
IBN has been the Vehicle Network Management with its distributed and
self-organizing networks composed of diverse and enormous number of high-speed
vehicles; Here, IBN-based policies are expected to manage vehicular networks towards
an intelligent network behavior according to the underlying applications or IoT
workflows, as well as help in minimizing the energy consumption and reducing of
network latency. An example can be seen in \cite{singh2020intent}, where
the authors designed an IBN control framework over the SDN architecture in the vehicular
edge computing ecosystem, achieving the set goals of improving on network energy efficient and latency
reduction. network latency. High-level programming abstractions and declarative
languages \cite{subramanya2016intent} are posed to significantly benefit
mobile networks and applications as well, due  to the increase in their
complexity over the years. This can be seen in the design and prototype
implementation of an Intent-based mobile backhauling interface for 5G networks
presented in \cite{subramanya2016intent} and in \cite{amrutha2015voice}, where
a wireless home automation system with automatic spoken word into text commands
translation was proposed.

The possibilities of smart network management and control with IBNs have also been appealing
with industrial solutions, with Huawei \cite{huawei2019} developing an IBN
product called Intent-Driven Network (IDN) for its cloud solution. Likewise
OpenDayLight Controller Platform has been involved in numerous strategies on
enabling the utilization of intents with Network Intent Composition
(NIC) \cite{nic2016}, Group-based  Policy (GBP) intent abstraction \cite{gbp2016},
and NEMO \cite{nemo2016} language.

The full potential of IBN approach is probably most evident in network security,
where the security can be incorporated as an integral part of
intent-based networking across “intent,” “automation,” and “assurance”
functions. With their continuous loop frameworks, IBNs have proven to provide
ongoing protection and alignment to security policy and compliance
requirements \cite{laliberte2018journey}. Moreover, IBN
implementation can provide traffic segmentation based on business roles for
connections to remote or cloud computing sites to simplify threat protection and
reduce the attack surface \cite{laliberte2018journey}. A policy-driven security orchestration framework were presented in
has also been presented in \cite{zarca2020semantic} in SDN/NFV-aware IoT
scenarios.

%\hl{\textit{Contributions of this paper}}\\

\vspace{-0.2 cm}
\section{System Design for ICT Supply Chains}

\begin{figure*}[!t]
    \centering
    \includegraphics[width=0.85\textwidth]{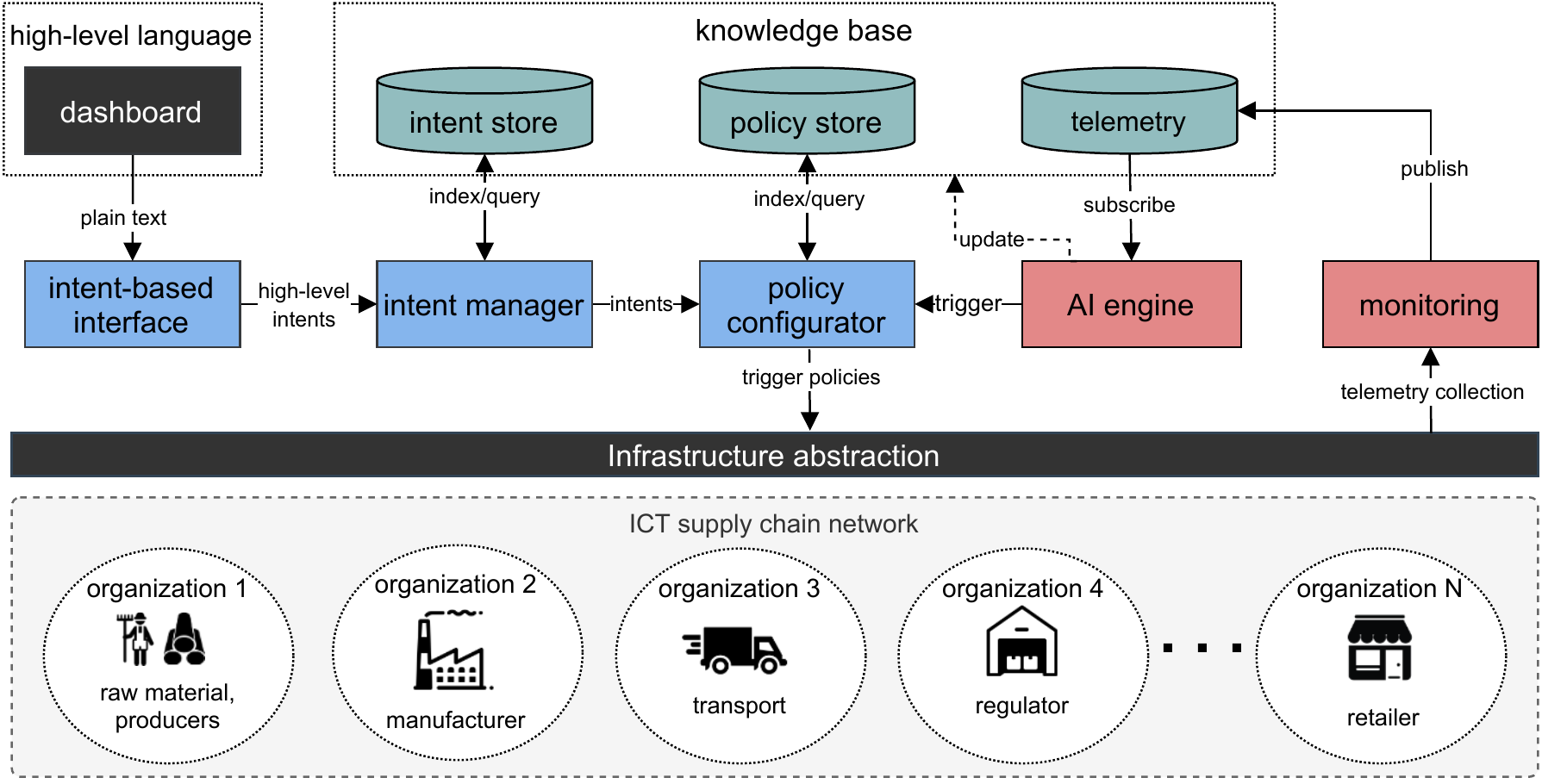}
    \caption{Intent-based translator design for ICT supply chain networks}
    \label{fig:arch}
\end{figure*}
\subsection{Requirements Engineering}

%\section{Discussion and Future Direction}

Deployment of IBN network orchestration and policy-based automation solutions in
ICT supply chain management carries a great potential. But, in order to leverage
this potential the future system design  needs to adequately address some of the
main requirements which affect both of these domains, and can be assumed to be
critical in their integration.

\textbf{Centralized/decentralized orchestrating approach:}
\noindent When opting for a straightforward automation and deployment of
intents, with the total control being handed over to a single node with the
complete knowledge of the network, the choice would be a centralized network
orchestrator. This approach comes at a price of that the central node machine
running the orchestrator would have to have significant storage and processing
power in order to maintain information for potentially very large network.  With
a decentralized approach, distributed orchestrators maintain an incomplete
information of the network, allowing higher levels of scalability but
introducing its own challenge regarding the efficient processing of intents
across multiple orchestrating nodes \cite{Shry}. A similar discussion is led in
assessing the  advantages and disadvantages of centralized and decentralized
control and orchestration strategies in todays complex ICT supply chains
environments has to be considered during the design phase of the management
system. The more traditional centralized approaches which often assume complete
information sharing among different SC organizational system components with the
a top hierarchical decision making authority have been noted to offer more
stable environments, faster decision making and easier coordination. On the
other side decentralized approach, with orchestration and decision making left
to individual SC organizational system components, benefits the SC efficiency.

\textbf{Intent specification:}
One of the crucial issues in IBN has been on how to actually specify high-level
intents, and here there is still a lack of standardized solutions. And while
there are ongoing efforts towards improving the ways intents can be described
and approaching the naturally spoken languages (e.g NEMO and INDIRA), how this
will relate to translating intent-based policies is to be
discussed, but a general language model will be required in order to
provide a scalable solution for different ICT supply chain systems.

\textbf{AI technologies:}
Both IBN and ICT supply chains have been greatly influenced by the advances in
the domains of AI and ML. In case of AI-powered IBNs, the performances have been
improved by using AI in network analysis and intent implementation as well as
using AI-based heuristics for solving constrained optimization problems. Here
the challenge lies in a critical issue for any AI/ML algorithm - the lack of
data. To tackle this we can assume that IBN will have to operate for certain
amount of time to have enough data available, so it will be necessary to assess
how to use data collected from the pre-IBN deployment.  Moreover, when
leveraging the potential of AI techniques for IBN applications in supply chain
management (SCM) there are two additional questions that need to be addressed.
Firstly, the question of leveraging the potential of natural language processing
(NLP) \cite{TOORAJIPOUR2021502} in improving human–machine interactions in SCM
should be addressed. Secondly, as supply chains are particularly important from
industrial networks perspective, in case of an incident it is a requirement to
obtain the explanation behind it, which can be difficult in case it is a
consequence of an IBN automated algorithm reaction \cite{saha2018intent}.

\textbf{Closed-loop verification:}
With closed-loop verification method, IBNs maintain the enforced network state
and configuration through intents, which can be expressed as series of smaller
network element configurations \cite{8968429}. The problem lies in the
complexity of this process, which assumes constant system and network
configuration adjustments in order to satisfy desired system requirements and
can result in autonomous remediation errors or unwanted delays. The complex
system architectures as the ones behind SC scenarios will need to improve on
this method, as a seemingly small configuration adjustments can affect the
entire network across.

\textbf{Single vs Multi-Domain:}
Majority of the IBN deployment and research work focuses on single-domain
network approach, with the orchestrator centralization. However, this kind of
approach is insufficient in addressing the SCM scenarios, as they normally
extend across multiple companies, decision-makers and network domains. To
address this particular issue there is a need for shifting towards a more
multi-domain oriented IBN approach, where an intent framework will be
implemented in a manner that would ensure end-to-end network orchestration
across multiple potential SC domains.

\subsection{Architecture Design}

The intent based translator aims to automate the configuration management of ICT
supply chain networks. Thus, we propose a general software architecture for the
intent abstraction layer, as shown in Fig.\ref{fig:arch}. Supply chains
include various organizational partners that own different parts of the supply
chain and need to cooperate in order to successfully deliver a final product or
service. The management solution of such network will allow different network
administrators to easily configure and control the network via high-level
intents.  A plain text, that is restricted by a predefined high-level language,
is input by the user, using  a dashboard  that communicates the information via
an intent-based interface. This interface is developed as to
facilitate the human machine interactions, considering the ability to control
the infrastructure using high-level intents for security assessment, network
service management and control. From here, as shown in Fig.\ref{fig:arch}, high-level
intents are translated and parsed by the intent manager into a structured format, using natural language
processing techniques. A knowledge base is then used to store the intent vocabularies
in a json format, where it is used by the intent manager to determine the intent
specifications. The vocabularies inside the knowledge base can be
enriched dynamically, improving the system scalability and capability to manage supply
chains of different contexts. After producing a structured format of an intent
with user requirement specifications, the intent is sent to a so-called policy
configurator component, responsible on matching the user requirements with
existing policies that can be applied in the infrastructure. This component will then
check the existing policies stored in the knowledge base in
a defined policy store that can be updated and accessed by developers and
experts in order to define new policies. The created policies are mapped by the
policy configurator into a lower level format, after checking for possible conflicts
among them. The policy configurator is also responsible for triggering the network
controller, using  defined APIs. In proposed system we also consider the possibility of
smartly reconfiguring the network based on insights from the system. Metrics are
collected by monitoring tools from all the network elements, and used to periodically
update the knowledge base, which includes telemetry as well as storing policies
and intent vocabularies. A component called AI Engine is responsible on
analyzing the data coming from the system, predicting the network status, and
automatically configuring the policies.
\vspace{-0.2 cm}
\subsection{Example of Supply Chain Assets Access Control}
\begin{figure}[!t]
    \centering
    \includegraphics[width=1\columnwidth]{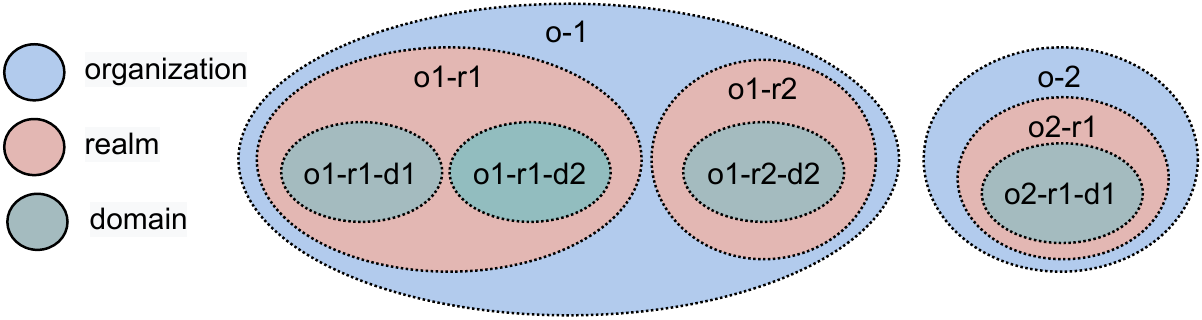}
    \caption{Supply chain abstraction building block}
    \label{fig:sc-org}
\end{figure}

An overview of the SC abstraction proposed in our example is illustrated in Fig.
\ref{fig:sc-org}. Organizations are in the top of the hierarchy and represent
companies, consortiums or institutions that belongs to the supply chain. Realms
consist of an environment within an organization that shares common security
policies and rules. Domains are a set of assets with certain relationship, for
instance, they share the same network, location or infrastructure.

\subsubsection{Intent Language Definition}

We define a simple grammar for SC asset access control as an intent language,
based on a context-free grammar standard: the Extended  Backus-Naur  Form (EBNF)
\cite{scowen1993generic}. The defined grammar in Grammar \ref{gra:sc} is a CNL
that allows the system administrator to grant access to users for a certain
asset in the system, restricted to specific organization, realm, or domain at a
certain working time. This proposal is a general definition of the grammar, and
can be enriched with more details related to the nature of an asset, the
composition of a supply chain, the structure of a company, and finally the type
of users.
\begin{Grammar}
    \begin{grammar}

        <intent> ::= <users> is <permission> to access to <asset> <spot>  at <timeframes>

        <users> ::= <user> (, and ) <user>

        <user> ::= [a-z0-9]+

        <permission> ::= allowed | blocked

        <asset> ::= domain | realm | organization

        <spot> ::= [a-z0-9]+

        <timeframes> ::= <timeframe> (, and ) <timeframe>

        <timeframe> ::= <morning shift> | <late shift> | <night shift>

        <morning shift> ::= [6:00-13:59]

        <late shift> ::= [14:00-21:59]

        <night shift> ::= [22:00-5:59]

    \end{grammar}
    \caption{SC Assets Access Control Intent}\label{gra:sc}
\end{Grammar}
%\vspace{-4mm}
\subsubsection{Intent Manager}
It parses the input plain text into a json file using the
defined high-level language and regular expressions (regex approach). The
parsing process can be improved using ML techniques to allow the user to employ
multiple vocabulary options. The intent manager reads the intent store from the knowledge base and try to match the
existing vocabularies and structures with the input text. Finally a json file
with user specifications is generated and sent to the policy configurator.

\subsubsection{Policy Configurator}
It matches the requirements specifications received from
the intent manager with the existing policies stored in the policy store. It solves conflict between intents and generates valid  policies.
Conflicts can happen between policies while compiling intent specifications.
A user can be restricted from access to assets in realm \emph{o1-r1} using a
following intent : \emph{user-x is not allowed to access to realm o1-r1}.
The policy configurator will map  the intent into high level policies such as:\\ \indent\indent \emph{check user-x in database of Users\\\indent\indent block user-x to access assets in o1-r1 \\ \indent\indent alert admin in o1}\\
When the administrator wants to give the user
access to assets in organization \emph{o1}, the following intent can be
written: \emph{user-x is allowed to access to organization o1}.
This intent will create a conflict with the previous intent during the policy translation,
as it will allow the user to access to realm \emph{o1-r1}, which is inside the organization \emph{o1}.
In order to solve this issue, conflict checking needs to be considered and implemented.
A solution to the mentioned conflict can be translated to the following high-level
policies: \\ \indent\indent \emph{check user-x in database of Users\\\indent\indent allow user-x to access assets in o1 except o1-r1 \\ \indent\indent alert admin in o1}\\ Other conflicts should be analysed considering the organizational structure, the existing policies and the possible human intervention.
\subsubsection{Knowledge Base}
It consists of three main components: intent store, policy
store, and telemetry. The intent store is used to store a dictionary of
vocabularies and options that are required to parse intents. The policy store
represents the policy graphs defined by experts, in our case the policy graph
represent the access to assets based on the organizational hierarchy. And
finally the telemetry is used to store the needed metrics to detect anomalies in
the system.

\subsubsection{AI Engine}
NLP techniques, such as long-short time memory (LSTM), can be implemented to
enhance the proposed CNL and provide more flexibility to the user while writing
the intents. The parsing of the input plain text considers a rich dictionary of
vocabularies and expressions that can mean the same input specifications. AI
techniques can also be used to enhance the policy configuration by solving
policy conflicts. Metrics collected from system and stored in the knowledge base
can be used to predict if a user has to be blocked or not based on his monitored
behavior, and ML techniques are a great candidate for such task of pattern
recognition and security assessment.

\section{Conclusion}
This  paper  proposes a general design of an intent-based translation system
that manages ICT supply chain networks, considering AI enabling technologies to
ensure a valid translation of user requirements, network security and
scalability. With our system design we intend to move the complexity of network management, which requires very high level of expertise from the system administrator by automating policy building process through AI and ML techniques. We analysed an use case example of an ICT supply chain management task  - access control policy configuration using high level intents. Also, we discussed
the engineering requirements and challenges in integrating IBN approach with ICT
supply chains, in terms of network orchestration complexity, the application of
AI techniques, and the ambiguity of defining high level intents for ICT supply
chain policies.

\section*{Acknowledgment}
This work is  partially funded by European Commission under the H2020-952644 contract for project FISHY: A coordinated framework for cyber resilient supply chain systems over complex ICT infrastructures.

\bibliographystyle{IEEEtran}
\bibliography{mybib.bib}

\end{document}